\documentclass[slac_one]{revtex4}
\usepackage{graphicx}
\usepackage{fancyhdr}
\begin{document}

\title{{\small{2005 International Linear Collider Workshop - Stanford,
U.S.A.}}\\ 
\vspace{12pt}
Probing Electroweak Top Quark Couplings at Hadron and Lepton Colliders} 

%

\author{U.~Baur}
\affiliation{Physics Department, State University of New York at
Buffalo, Buffalo, NY 14260, USA}

\begin{abstract}
The International Linear Collider (ILC) will be able to precisely measure the
electroweak couplings of the top in $e^+e^-\to t\bar t$. We compare the
limits which can be achieved at the ILC with those which can be obtained
in $t\bar 
t\gamma$ and $t\bar tZ$ production at the Large Hadron Collider (LHC). 
\end{abstract}

\maketitle

\thispagestyle{fancy}


\section{INTRODUCTION} 

Although the top quark was discovered almost ten years
ago~\cite{topcdf,topd0}, many of its properties are still only poorly
known~\cite{Chakraborty:2003iw}.  In particular, the couplings of the
top quark to the electroweak (EW) gauge bosons have not yet been
directly measured.  
Current data provide only weak constraints on the couplings of the top
quark with the EW gauge bosons, except for the $ttZ$ vector and axial
vector couplings which are rather tightly but indirectly constrained
by LEP data; and the right-handed $tbW$
coupling, which is severely bound by the observed $b\to s\gamma$
rate~\cite{Larios:1999au}.  

At an $e^+e^-$ linear collider with $\sqrt{s}=500$~GeV and an
integrated luminosity of $100-200$~fb$^{-1}$ one can hope to measure
the $ttV$ ($V=\gamma,\,Z$) couplings in top pair production with a few-percent
precision~\cite{Abe:2001nq}.  However, the process
$e^+e^-\to\gamma^*/Z\to t\bar{t}$ is sensitive to both $tt\gamma$
and $ttZ$ couplings and significant cancellations between the various
couplings can occur.  At hadron colliders, $t\bar{t}$ production is so
dominated by the QCD processes $q\bar{q}\to g^*\to t\bar{t}$ and
$gg\to t\bar{t}$ that a measurement of the $tt\gamma$ and $ttZ$
couplings via $q\bar{q}\to\gamma^*/Z^*\to t\bar{t}$ is hopeless.
Instead, the $ttV$ couplings can be measured in QCD $t\bar{t}\gamma$
production, radiative top quark decays in $t\bar{t}$ events
($t\bar{t}\to\gamma W^+W^- b\bar{b}$), and QCD $t\bar{t}Z$
production~\cite{Baur:2004uw}. $t\bar{t}\gamma$ production and radiative
top quark decays 
are sensitive only to the $tt\gamma$ couplings, whereas $t\bar{t}Z$
production gives information only on the structure of the $ttZ$
vertex.  This obviates having to disentangle potential cancellations
between the different couplings.

In this contribution we briefly review the measurement of the $ttV$
couplings at the LHC and compare the expected sensitivities with the
bounds which one hopes to achieve at an $e^+e^-$ linear collider. 

\section{General \boldmath{$ttV$} Couplings}

The most general Lorentz-invariant vertex function describing the
interaction of a neutral vector boson $V$ with two top quarks can be
written in terms of ten form factors~\cite{Hollik:1998vz}, which are
functions of the kinematic invariants.  In the low energy limit,
these correspond to couplings which multiply dimension-four or -five 
operators in an effective Lagrangian, and may be complex.  If $V$ is 
on-shell, or if $V$ couples to effectively massless fermions, the 
number of independent form factors is reduced to eight.  If, in 
addition, both top quarks are on-shell, the number is further reduced 
to four.  In this case, the $ttV$ vertex can be written in the form
\begin{equation}\label{eq:anomvertex}
\Gamma_\mu^{ttV}(k^2,\,q,\,\bar{q}) = -ie \left\{
  \gamma_\mu \, \left( F_{1V}^V(k^2) + \gamma_5F_{1A}^V(k^2) \right)
+ \frac{\sigma_{\mu\nu}}{2m_t}~(q+\bar{q})^\nu 
   \left( iF_{2V}^V(k^2) + \gamma_5F_{2A}^V(k^2) \right)
\right\} \, ,
\end{equation}
where $e$ is the proton charge, 
$m_t$ is the top quark mass, $q~(\bar{q})$ is the outgoing top
(anti-top) quark four-momentum, and $k^2=(q+\bar{q})^2$.  The terms
$F_{1V}^V(0)$ and $F_{1A}^V(0)$ in the low energy limit are the $ttV$ 
vector and axial vector form factors.  The coefficients 
$F_{2V}^\gamma(0)$ and $F_{2A}^\gamma(0)$ are related to the magnetic 
and ($CP$-violating) electric dipole form factors.

In $t\bar{t}V$ production, one of the top quarks coupling to $V$ is
off-shell.  The most general vertex function relevant for $t\bar{t}V$
production thus contains additional couplings, not included in
Eq.~(\ref{eq:anomvertex}). These additional couplings are irrelevant
in $e^+e^-\to t\bar{t}$, where both top quarks are on-shell. 

In $e^+e^-\to t\bar{t}$ one often uses the following parameterization for
the $ttV$ vertex:
\begin{equation}\label{eq:gordon}
\Gamma_\mu^{ttV}(k^2,\,q,\,\bar{q}) = ie \left\{
  \gamma_\mu \, \left(  \widetilde{F}_{1V}^V(k^2)
                      + \gamma_5\widetilde{F}_{1A}^V(k^2) \right)
+ \frac{(q-\bar{q})_\mu}{2m_t}
    \left(  \widetilde{F}_{2V}^V(k^2)
          + \gamma_5\widetilde F_{2A}^V(k^2) \right)
\right\} .
\end{equation}
Using the Gordon decomposition, it is easy to show that
Eqs.~(\ref{eq:anomvertex}) and~(\ref{eq:gordon}) are equivalent for
onshell top quarks and that the form
factors $\widetilde F^V_{iV,A}$ and $F^V_{iV,A}$ ($i=1,\,2$) are
related by
\begin{equation}
\label{eq:rel1}
\widetilde F^V_{1V} = -\left( F^V_{1V}+F^V_{2V} \right) \, , \qquad
\widetilde F^V_{2V}  =  \phantom{-} F^V_{2V} \, , \qquad
\widetilde F^V_{1A} = -F^V_{1A} \, , \qquad
\widetilde F^V_{2A} =  -iF^V_{2A} \, .
\label{eq:rel4}
\end{equation}

\section{\boldmath{${t\bar{t}\gamma}$} Production at the LHC}

The process $pp\to t\bar{t}\gamma$
followed by $t\to Wb$ leads either to a
$\gamma\ell\nu_\ell\ell'\nu_{\ell'}b\bar{b}$ final state if both $W$
bosons decay leptonically, to a $\gamma\ell\nu_\ell b\bar{b}jj$ final
state if one $W$ decays leptonically and the other decays hadronically, 
or to a $\gamma b\bar{b}+4j$ final state if both $W$ bosons decay
hadronically.  The $\gamma b\bar{b}+4j$ final state has the largest
BR.  However, it is plagued by a large QCD background, so we ignore
it.  The dilepton final state, although less contaminated by
background, has a BR about a factor~6 smaller than that of the
so-called lepton+jets mode.  In the following, we therefore
concentrate on $pp\to \gamma\ell\nu_\ell b\bar{b}jj$
with $\ell=e\,,\mu$.  We assume that both
$b$ quarks are tagged with a combined efficiency of $\epsilon_b^2=40\%$.

We perform our calculation for general $tt\gamma$ couplings of the
form of Eq.~(\ref{eq:anomvertex}). We otherwise assume the SM to be
valid. Our calculation includes top quark and $W$ decays with full spin
correlations and finite width effects.  All resonant Feynman diagrams
contributing to the lepton+jets final state are included, i.e. besides
$t\bar{t}\gamma$ production, we automatically take into account 
top quark pair production where one of the top quarks decays 
radiatively, $t\to Wb\gamma$.  

We impose standard acceptance cuts for leptons, jets, and the missing
transverse momentum. A detailed list can be found in
Ref.~\cite{Baur:2004uw}. We also include minimal detector effects via Gaussian
smearing of parton momenta according to CMS~\cite{cms} expectations, and
take into account the $b$ jet energy loss via a parameterized function.
Since we are interested in photon emission from top quarks, we would
like to suppress radiation from $W$ decay products, as well as
emission from $b$ quarks. Imposing a large
separation cut of $\Delta R(\gamma,b)>1$ reduces photon
radiation from the $b$ quarks.  Photon emission from $W$ decay
products can essentially be eliminated by requiring that
$m(jj\gamma) > 90~{\rm GeV}$ and $
m_T(\ell\gamma;p\llap/_T) > 90~{\rm GeV,}$
where $m(jj\gamma)$ is the invariant mass of the $jj\gamma$ system, and
$m_T(\ell\gamma;p\llap/_T)$ is the $\ell\gamma p\llap/_T$ cluster
transverse mass, which peaks sharply at $m_W$. In addition we require
that the event is consistent either 
with $t\bar{t}\gamma$ production, or with $t\bar{t}$ production with
radiative top decay.  This will reduce the singly-resonant and 
non-resonant backgrounds. The invariant mass and cluster transverse mass
cuts which are imposed to accomplish this can be found in
Ref.~\cite{Baur:2004uw}. 

The most important irreducible background processes that remain after 
imposing the cuts described above, are single-top processes
($(t\bar{b}\gamma + \bar{t}b\gamma)+X$), and the
non-resonant process $pp \to
W(\to\ell\nu)\gamma b\bar{b}jj$.  We calculate the irreducible
background processes at leading order in QCD including the full set of
contributing Feynman diagrams using {\tt
MADEVENT}~\cite{Maltoni:2002qb}. 
The potentially most dangerous reducible background is $t\bar{t}j$ 
production where one of the jets in the final state fakes a photon.

\begin{figure}
\begin{center}
\begin{tabular}{cc}
\includegraphics[width=8.7cm]{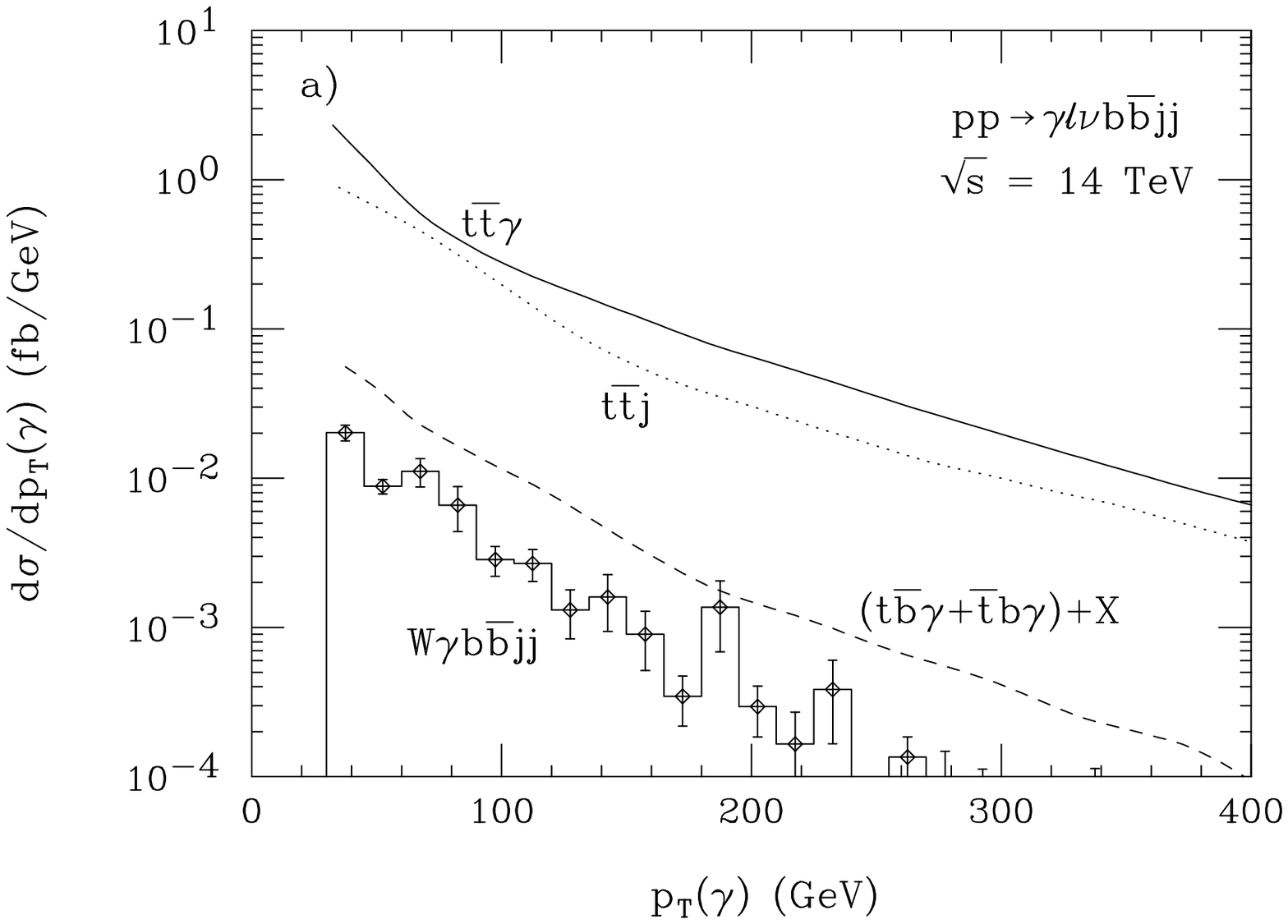} &
\includegraphics[width=8.7cm]{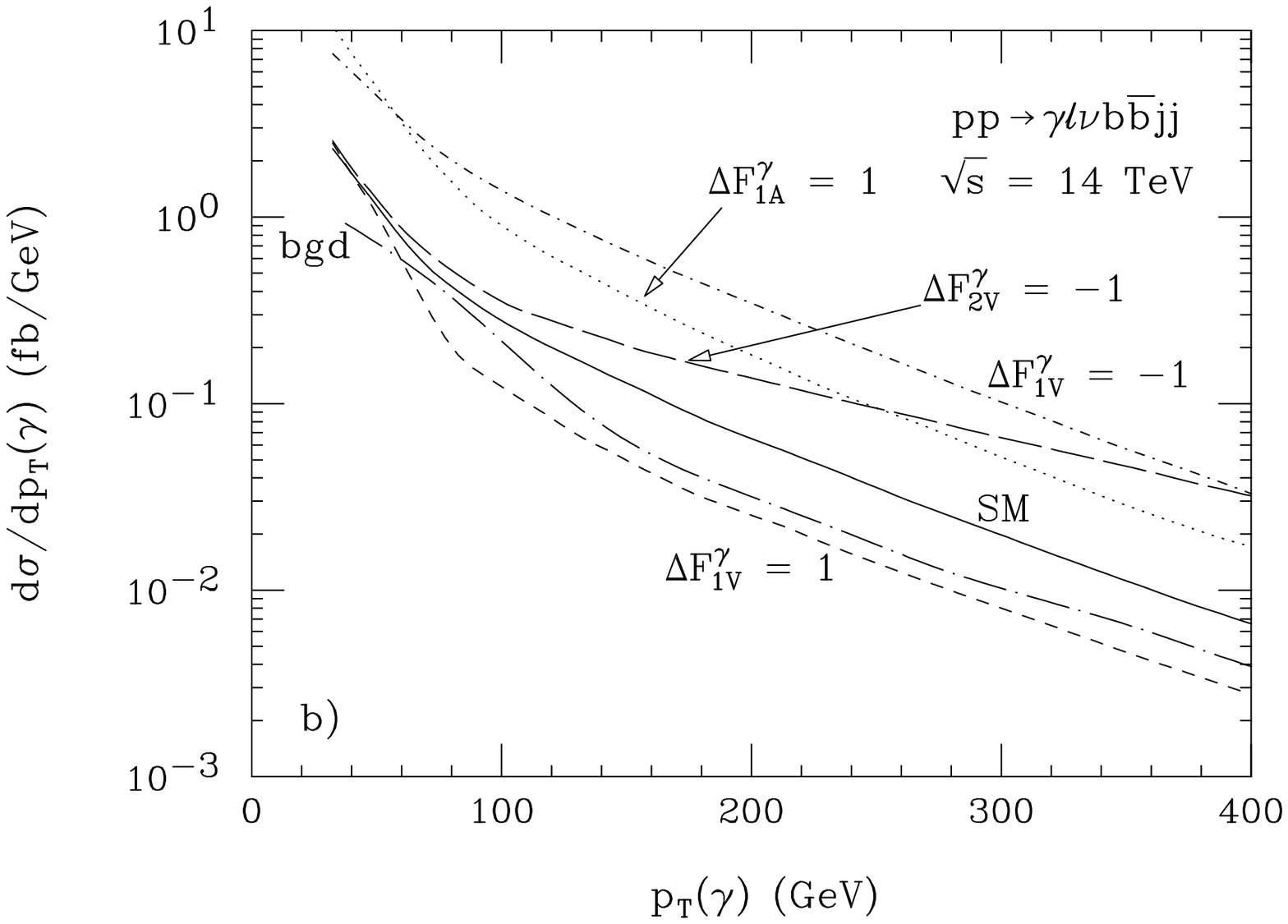}
\end{tabular}
\end{center}
\caption{\label{fig:fig1}{The differential cross sections as a function
of the photon  transverse momentum for $\gamma\ell\nu_\ell b\bar{b}jj$
production at the LHC. Part a) shows the SM signal and the various
contributions to the background. Part b) shows the SM signal and
background, and the signal for various anomalous $tt\gamma$ couplings.}}
\end{figure}
In Fig.~\ref{fig:fig1}a we show the photon transverse momentum
distributions of the $t\bar{t}\gamma$ signal and the backgrounds
discussed above. The $t\bar{t}j$ background is seen to
be a factor~2 to~3 smaller than the $t\bar{t}\gamma$ signal for the
jet -- photon misidentification probability (
$P_{j\to\gamma}=1/1600$~\cite{atlas_tdr}) used. The 
$(t\bar{b}\gamma +
\bar{t}b\gamma)+X$ and $W\gamma b\bar{b}jj$ backgrounds both are found
to be more than an order of magnitude smaller than the $t\bar{t}j$
background.

The photon transverse momentum distributions in the SM and for various
anomalous $tt\gamma$ couplings, 
together with the $p_T(\gamma)$ distribution of the
background, are shown in
Fig.~\ref{fig:fig1}b. Only one coupling at a time is allowed to deviate
from its SM prediction. 

\section{\boldmath{${t\bar{t}Z}$} Production at the LHC}

The process $pp \to t\bar{t}Z$ leads
to either ${\ell'}^+{\ell'}^-\ell\nu b\bar{b}jj$ or
${\ell'}^+{\ell'}^- b\bar{b}+4j$ final states if the $Z$-boson
decays leptonically and one or both of the $W$ bosons decay
hadronically. If the $Z$ boson decays into neutrinos and both $W$ bosons
decay hadronically, the final state consists of
$p\llap/_Tb\bar{b}+4j$. Since there is essentially no phase space for
$t\to WZb$ decays ($BR(t\to WZb)\approx 3\cdot
10^{-6}$~\cite{Mahlon:1994us}), these final states
arise only from $t\bar tZ$ production. 

In order to identify leptons, $b$ quarks, light jets and the missing
transverse momentum in dilepton and trilepton events, we impose the same
cuts as for $t\bar t\gamma$ production. We also require that there is a
same-flavor, opposite-sign lepton pair with 
invariant mass near the $Z$ resonance, $m_Z - 10~{\rm GeV} < m(\ell\ell)
< m_Z + 10~{\rm GeV}$. 

The main backgrounds contributing to the trilepton final state are
singly-resonant $(t\bar{b}Z+\bar{t}bZ)+X$ ($t\bar{b}Zjj$,
$\bar{t}bZjj$, $t\bar{b}Z\ell\nu$ and $\bar{t}bZ\ell\nu$) and
non-resonant $WZb\bar{b}jj$ production. In the dilepton case, the main
background arises from 
$Zb\bar{b}+4j$ production, which we calculate using {\tt
ALPGEN}~\cite{Mangano:2002ea}.  To adequately suppress it, we
additionally require that events have at least one combination of jets
and $b$ quarks which is consistent with the $b\bar b+4j$ system
originating from a $t\bar t$ system. Once these cuts have been imposed,
the $Zb\bar{b}+4j$ background is important only for $p_T(Z)<100$~GeV. 

\begin{figure}
\begin{center}
\begin{tabular}{cc}
\includegraphics[width=8.7cm]{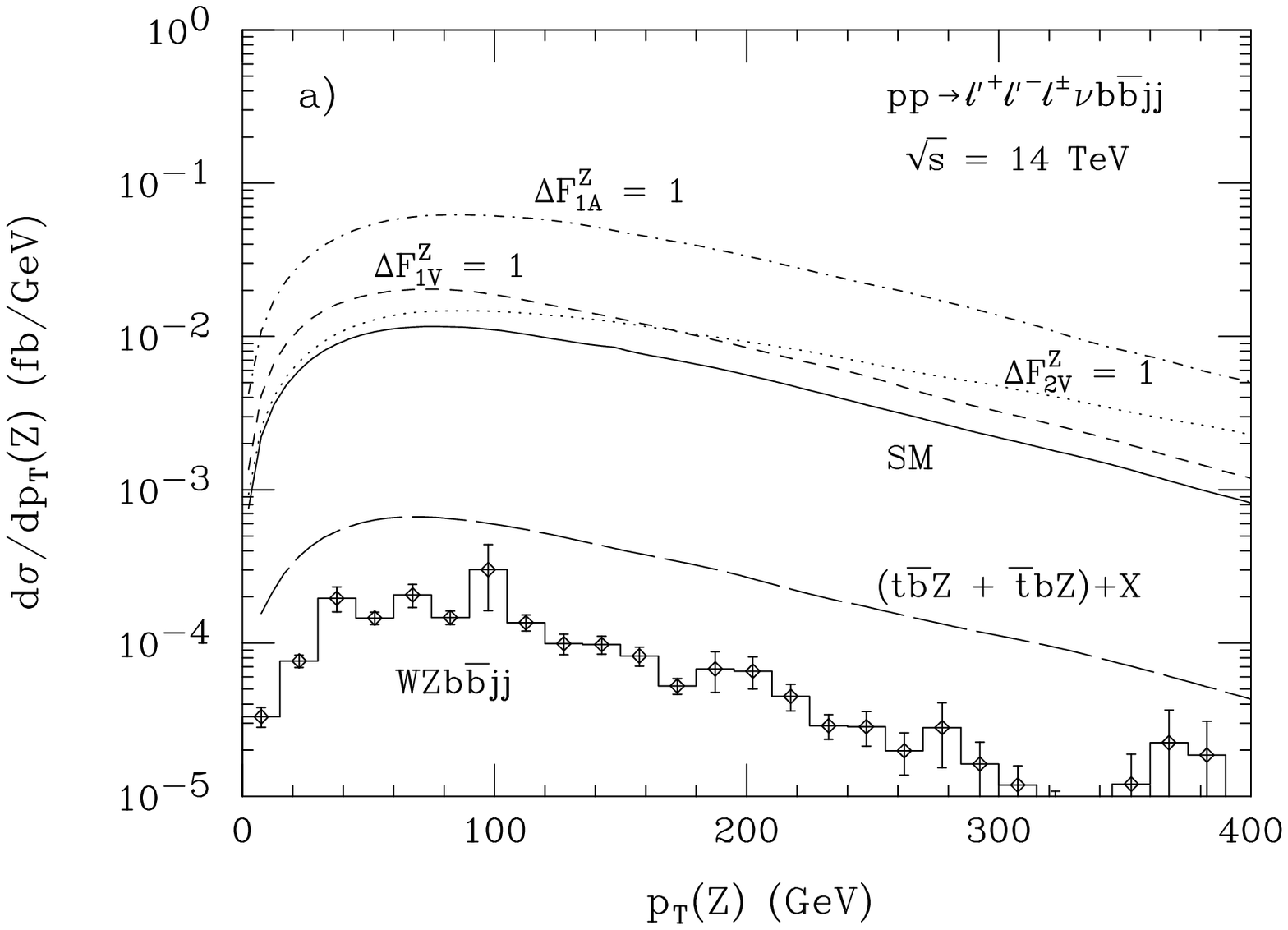} &
\includegraphics[width=8.7cm]{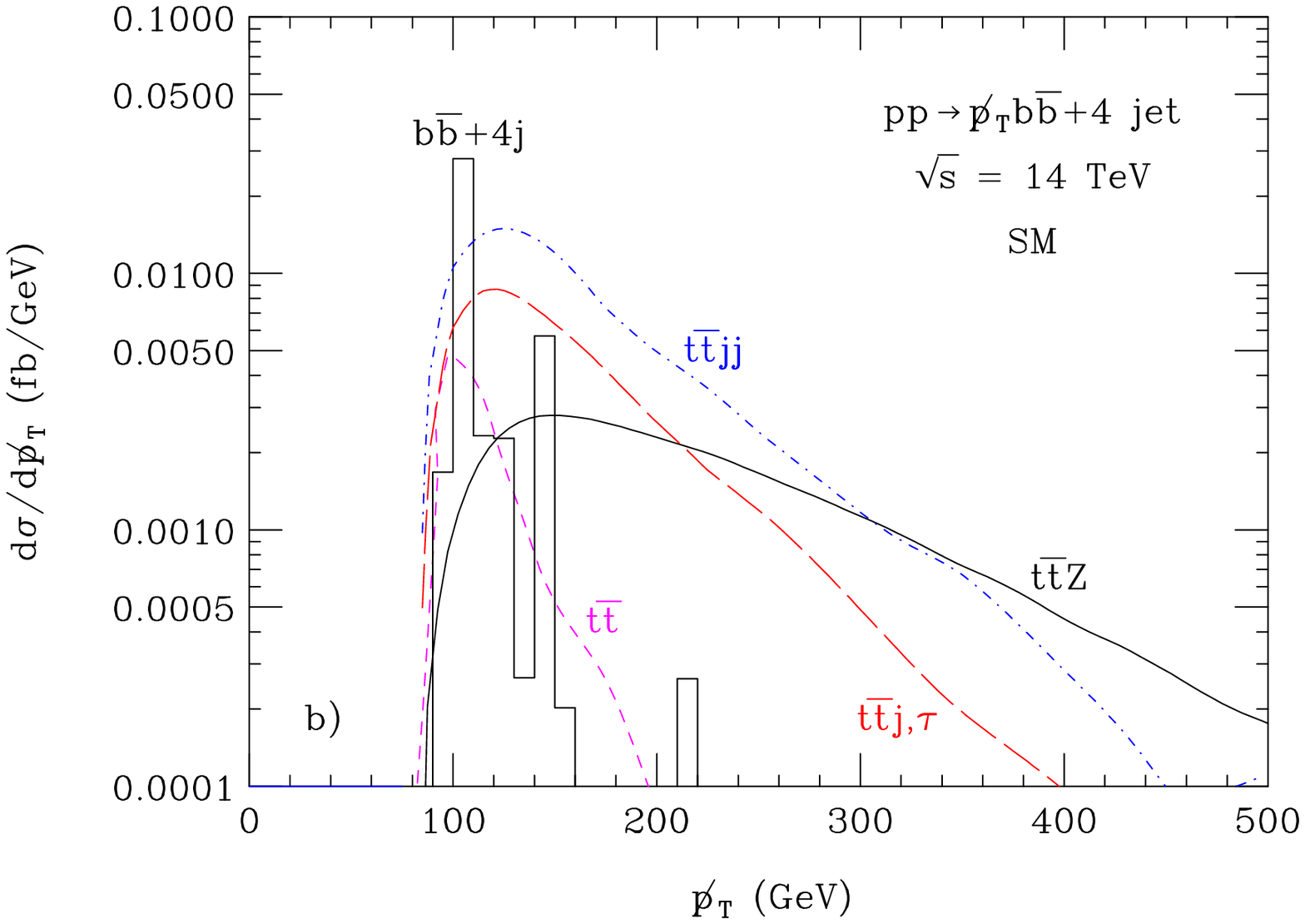}
\end{tabular}
\end{center}
\caption{\label{fig:fig2}{a) The differential cross sections at the LHC
as a function 
of $p_T(Z)$ for ${\ell'}^+{\ell'}^-\ell\nu b\bar{b}jj$ final states.  
Shown are the SM predictions for $t\bar{t}Z$ production, for several
non-standard $ttZ$ couplings, and for various
backgrounds.  Only one coupling at a time is allowed to deviate from 
its SM value. 
b) The differential cross sections as a function of the missing
transverse momentum for $p\llap/_Tb\bar b+4$j production at the
LHC.  Shown are the SM predictions for
$t\bar{t}Z$ production and for various backgrounds. }}
\end{figure}
The $Z$ boson transverse momentum distribution for the trilepton final
state is shown in
Fig.~\ref{fig:fig2}a for the SM signal and backgrounds, as well as for 
the signal with several non-standard $ttZ$ couplings. Only one coupling
at a time is allowed to deviate from its SM 
prediction.  The backgrounds are each more than one order of magnitude
smaller than the SM signal. Figure~\ref{fig:fig2}a shows that varying
$F^Z_{1V,A}$ leads mostly to a cross section 
normalization change, hardly affecting the shape of the $p_T(Z)$
distribution.  This is because, unlike in the $tt\gamma$ case, there
is no radiative top decay, i.e. no $t\bar{t}$ events where $t\to WZb$.
This implies that, for the cuts we impose, the $p_T(Z)$ distribution
for SM couplings and for $F^Z_{1V,A}=-F^{Z,SM}_{1V,A}$ are almost
degenerate. Currently, the SM $t\bar{t}Z$ cross section is known only at
LO, and 
has substantial factorization and renormalization scale uncertainty.
Since the backgrounds are insignificant, this normalization
uncertainty, and the sign degeneracy, will ultimately be the limiting
factors in extracting 
anomalous vector and axial vector $ttZ$ couplings.

For the $p\llap/_Tb\bar{b}+4j$~\cite{new} final state we require at
least 3~jets 
with $p_T>50$~GeV and $p\llap/_T>5~{\rm GeV}^{1/2}\sqrt{\sum p_T}$. The
largest backgrounds for this final state come from $t\bar t$ and $b\bar
b+4j$ production where one or several jets are badly mismeasured, from
$pp\to t\bar tjj$ with $t\bar 
t\to\ell^\pm\nu_\ell b\bar bjj$ and the charged lepton
being missed, and from $t\bar tj$ production, where one top decays
hadronically, $t\to Wb\to bjj$, and the other via $t\to Wb\to\tau\nu_\tau b$
with the $\tau$-lepton decaying hadronically, $\tau\to h\nu_\tau$. 

In Fig.~\ref{fig:fig2}b we show the missing transverse momentum
distributions of the SM $t\bar tZ\to p\llap/_Tb\bar{b}+4j$ signal (solid
curve) and various 
backgrounds. The most important backgrounds are $t\bar tjj$
and $t\bar tj$ production. However, the missing transverse momentum
distribution from these processes falls considerably faster than that of
the signal, and for $p\llap/_T>300$~GeV, the SM signal dominates. 

\section{Sensitivity Bounds for $ttV$ Couplings: LHC and ILC}

The shape and normalization changes of the photon or $Z$-boson
transverse momentum distribution can be used to derive
quantitative sensitivity bounds on the anomalous $tt\gamma$ and $ttZ$
couplings. For $t\bar{t}Z$ production, the $\Delta\Phi({\ell'}{\ell'})$
distribution provides additional information~\cite{Baur:2004uw}. In the
following we assume a normalization uncertainty of the SM cross
section of $\Delta{\cal N}=30\%$. 

Even for a modest integrated luminosity of 30~fb$^{-1}$, it will be 
possible to measure the
$tt\gamma$ vector and axial vector couplings, and the dipole form
factors, with a precision of typically $20\%$ and $35\%$,
respectively. For 300~fb$^{-1}$, the limits improve to $4-7\%$ for
$F^\gamma_{1V,A}$ and to about $20\%$ for $F^\gamma_{2V,A}$. At the
SLHC, assuming an integrated luminosity of 3000~fb$^{-1}$, one can
hope to achieve a $2-3\%$ measurement of the vector and axial vector
couplings, and a $10\%$ measurement of $F^\gamma_{2V,A}$, provided
that particle identification efficiencies are not substantially
smaller, and the reducible backgrounds not much larger, than what we
have assumed.

To extract bounds on the $ttZ$ couplings, we perform a simultaneous
fit to the $p_T(Z)$ and the $\Delta\Phi({\ell'}{\ell'})$
distributions for the trilepton and dilepton final states, and to the
$p\llap/_T$ distribution for the $p\llap/_Tb\bar b+4j$ final state. We
calculate sensitivity bounds for 
300~fb$^{-1}$ and 3000~fb$^{-1}$ at the LHC; for 30~fb$^{-1}$ the
number of events expected is too small to yield meaningful results.
For an integrated luminosity of 300~fb$^{-1}$, it will be possible to
measure the $ttZ$ axial vector coupling with a precision of $15-20\%$,
and $F^Z_{2V,A}$ with a precision of $50-55\%$.  At the SLHC, these
bounds can be improved by factors of about~1.6 ($F^Z_{2V,A}$)
and~3 ($F^Z_{1A}$).  The bounds which can be achieved for
$F^Z_{1V}$ are much weaker than those projected for $F^Z_{1A}$.  As
mentioned in Sec.~4, the $p_T(Z)$ distributions for the
SM and for $F^Z_{1V,A}=-F^{Z,SM}_{1V,A}$ are almost degenerate.
This is also the case for the $\Delta\Phi({\ell'}{\ell'})$
distribution.  In a fit to these two distributions, therefore, an area
centered at $\Delta F^Z_{1V,A}=-2F^{Z,SM}_{1V,A}$ remains which cannot
be excluded, even at the SLHC.  For $F^Z_{1V}$, the two regions merge,
resulting 
in rather poor limits.  The sensitivity bounds on $\Delta F^Z_{1A}$
improve by as much as a factor~2 if $\Delta{\cal N}$ can be reduced from
$30\%$ to $10\%$.  

\begin{table}[t]
\begin{center}
\caption{Sensitivities achievable at $68.3\%$ CL for the anomalous 
$ttV$ ($V=\gamma,\,Z$) couplings $\widetilde F^V_{1V,A}$ and
$\widetilde F^V_{2V,A}$ of Eq.~(\ref{eq:gordon}) at the LHC for
integrated luminosities of 300~fb$^{-1}$, and the ILC with
$\sqrt{s}=500$~GeV (taken from 
Ref.~\protect\cite{Abe:2001nq}).  Only one coupling at a time is
allowed to deviate from its SM value.\\}
\begin{tabular}{ccc|ccc}
\hline 
 coupling & LHC, 300~fb$^{-1}$ & $e^+e^-$~\protect\cite{Abe:2001nq} &
coupling & LHC, 300~fb$^{-1}$ & $e^+e^-$~\protect\cite{Abe:2001nq} \\
\hline
$\Delta\widetilde F^\gamma_{1V}$ & $\begin{matrix}{ +0.043 \\[-4pt]
-0.041}\end{matrix}$ & $\begin{matrix}{ +0.047 \\[-4pt]
-0.047}\end{matrix}$ , 200~fb$^{-1}$ & $\Delta\widetilde F^Z_{1V}$ &
$\begin{matrix} {+0.34 \\[-4pt] 
-0.72}\end{matrix}$ & $\begin{matrix} {+0.012 \\[-4pt]
-0.012}\end{matrix}$ , 200~fb$^{-1}$ 
\\
$\Delta\widetilde F^\gamma_{1A}$ & $\begin{matrix} {+0.051 \\[-4pt]
-0.048}\end{matrix}$ & $\begin{matrix} {+0.011 \\[-4pt]
-0.011}\end{matrix}$ , 100~fb$^{-1}$  & $\Delta\widetilde F^Z_{1A}$ &
$\begin{matrix} {+0.079 \\[-4pt] 
-0.091}\end{matrix}$ & $\begin{matrix} {+0.013 \\[-4pt]
-0.013}\end{matrix}$ , 100~fb$^{-1}$ 
\\
$\Delta\widetilde F^\gamma_{2V}$ & $\begin{matrix} {+0.038 \\[-4pt]
-0.035}\end{matrix}$ & $\begin{matrix}{ +0.038 \\[-4pt]
-0.038}\end{matrix}$ , 200~fb$^{-1}$  & $\Delta\widetilde F^Z_{2V}$ &
$\begin{matrix} {+0.26 \\[-4pt] 
-0.34}\end{matrix}$ & $\begin{matrix} {+0.009 \\[-4pt]
-0.009}\end{matrix}$ , 200~fb$^{-1}$ 
\\
$\Delta\widetilde F^\gamma_{2A}$ & $\begin{matrix} {+0.16 \\[-4pt]
-0.17}\end{matrix}$ & $\begin{matrix} {+0.014 \\[-4pt]
-0.014}\end{matrix}$ , 100~fb$^{-1}$   & $\Delta\widetilde F^Z_{2A}$ &
$\begin{matrix} {+0.35 \\[-4pt] 
-0.35}\end{matrix}$ & $\begin{matrix} {+0.052 \\[-4pt]
-0.052}\end{matrix}$ , 100~fb$^{-1}$  
\\     
\hline
\end{tabular}
\label{tab:tab1}
\end{center}
\end{table}
It is instructive to compare the bounds for anomalous $ttV$ couplings
achievable at the LHC with those projected for the ILC. The most
complete study of $t\bar{t}$ production at the ILC for general $ttV$
($V=\gamma,\,Z$) couplings so far is 
that of Ref.~\cite{Abe:2001nq}.  It uses the parameterization of
Eq.~(\ref{eq:gordon}) for the $ttV$ vertex function.  In order to
compare the bounds of Ref.~\cite{Abe:2001nq} with those anticipated
at the LHC, the limits on $F^V_{1V,A}$ and $F^V_{2V,A}$ have to be
converted into bounds on $\widetilde 
F^V_{1V,A}$ and $\widetilde F^V_{2V,A}$. Table~\ref{tab:tab1} compares
the bounds we obtain for $\widetilde F^V_{1V,A}$ and $\widetilde
F^V_{2V,A}$ with those reported for the ILC in Ref.~\cite{Abe:2001nq}. 
We show LHC limits only for an
integrated luminosity of 300~fb$^{-1}$.  The results of
Table~\ref{tab:tab1} demonstrate that the ILC, with the 
exception of $F^\gamma_{1V}$ and $F^\gamma_{2V}$, will be able to 
considerably improve the sensitivity limits which can be achieved at 
the LHC, in particular for the $ttZ$ couplings. 

\section{Conclusions}

The LHC will be able to perform first tests of the $ttV$ couplings. 
Already with an
integrated luminosity of 30~fb$^{-1}$, one can probe the $tt\gamma$ couplings
with a precision of about $10-35\%$ per experiment.  With higher
integrated luminosities one will be able to reach the few percent
region. The $t\bar{t}Z$ cross section with leptonic $Z$ decays is roughly a
factor~20 smaller than the $t\bar{t}\gamma$ rate.  It is therefore not
surprising that the sensitivity limits on the $ttZ$ couplings are
significantly weaker than those which one expects for the $tt\gamma$
couplings. 
The ILC, with the exception of $F^\gamma_{1V}$ and $F^\gamma_{2V}$, will
be able to further improve our knowledge of the $ttV$ couplings, in
particular in the $ttZ$ case.

{\sl This research was supported in
part by the National Science Foundation under grant No.~PHY-0139953.}


\end{document}